\documentclass[12pt]{article}

\usepackage{hyperref}
\usepackage[fleqn]{amsmath}
\usepackage{amssymb} 
\usepackage{bbm} 
\usepackage[small]{caption2} 
\usepackage{fleqn} 
\usepackage{graphicx} 
\usepackage{mathrsfs}   
\usepackage[small,loose]{subfigure}  
\usepackage{cite} 
\usepackage{color}
\usepackage{pst-node}
\usepackage{cancel}

\newcommand{\I}{\mathrm{i}}

\newcommand{\E}[1]{\ensuremath{\mathrm{E}_{#1}}} 

\newcommand{\SO}[1]{\ensuremath{\mathrm{SO}(#1)}}
\newcommand{\SU}[1]{\ensuremath{\mathrm{SU}(#1)}}
\newcommand{\U}[1]{\ensuremath{\mathrm{U}(#1)}}
\newcommand{\Z}[1]{\ensuremath{\mathbbm{Z}_{#1}}} 

\newcommand{\Id}[0]{\ensuremath{\mathbbm{1}}}

\numberwithin{equation}{section}
\numberwithin{table}{section}

\allowdisplaybreaks[1]

\usepackage{nicefrac}

\begin{document}

\begin{titlepage}

\vspace*{-3.0cm}
\begin{flushright}
\normalsize{DESY-12-047}\\
\normalsize{TUM-HEP 830/12}
\end{flushright}

\vspace*{1.0cm}

\begin{center}
{\Large\bf Origin of family symmetries}

\vspace{1cm}

\textbf{
Hans~Peter~Nilles$^{a}$,
Michael~Ratz$^{b}$,
Patrick~K.S.~Vaudrevange$^{c}$
}
\\[8mm]
\textit{$^a$\small~Bethe Center for Theoretical Physics and Physikalisches Institut der Universit\"at Bonn, Nussallee 12, 53115 Bonn, Germany}
\\[5mm]
\textit{$^b$\small~Physik--Department T30, Technische Universit\"at M\"unchen,
James--Franck--Stra\ss e, 85748 Garching, Germany}
\\[5mm]
\textit{$^c$\small~Deutsches Elektronen--Synchrotron DESY, Notkestra\ss e 85, 22607 Hamburg, Germany}
\end{center}

\vspace{1cm}

\vspace*{1.0cm}

\begin{abstract}
Discrete (family) symmetries might play an important role in models 
of elementary particle physics. We discuss the origin of such
symmetries in the framework of consistent ultraviolet
completions of the standard model in field and string theory.
The symmetries can arise due to special geometrical properties 
of extra compact dimensions and the localization of fields in
this geometrical landscape. We also comment on  anomaly 
constraints for discrete symmetries.
\end{abstract}

\end{titlepage}

\newpage

\section{Introduction}

Discrete symmetries play an important role in particle physics. Apart from the
fundamental space--time symmetries $P$, $C$ and $T$, there are various well
known examples such as the so--called matter or $R$ parity in the minimal 
supersymmetric standard model (MSSM).  There are good reasons for using
discrete rather than continuous symmetries. Models with spontaneously broken
global continuous symmetries exhibit Goldstone bosons which are
typically  phenomenologically unacceptable.  Moreover, there are strong
arguments that a continuous symmetry has either to be gauged or it will be
broken by quantum gravity effects (see e.g.\ \cite{Banks:2010zn} for a recent
discussion). In contrast to the fundamental symmetries, discrete symmetries 
are often just imposed by hand for phenomenological reasons.  While
introducing such symmetries can be a useful tool in bottom--up model building
it appears worthwhile to clarify the origin of a given symmetry. Given a
deeper understanding of how such symmetries arise, one might be able to obtain
a more fundamental understanding of observations, such as the repetition of
families and the flavor structure.

Discrete symmetries come in various classes.  Various generation--dependent
flavor symmetries, have been proposed in order to explain the pattern of quark 
and lepton Yukawa couplings, and to control
higher--dimensional operators (see e.g.\ \cite{Ishimori:2010au} for
a quite recent review and other contributions of this
special issue \cite{SpecialIssue:2012x} for more references). Apart from these there are
generation--independent symmetries, introduced in order to cure certain
shortcomings of extensions of the standard model such as the MSSM. For 
example, dangerous proton decay operators are forbidden by matter parity 
\cite{Fayet:1977yc,Dimopoulos:1981dw}, baryon triality \cite{Ibanez:1991pr}, 
proton hexality \cite{Dreiner:2005rd} and $\Z{4}^R$ \cite{Lee:2010gv}.
Further, discrete symmetries of high order can manifest themselves as accidental
global $\U{1}$ symmetries in the (truncated) low--energy effective theory. Such
accidental symmetries can be used for example in two ways: as (anomalous)
Peccei--Quinn symmetry addressing the strong CP problem (cf.\ the discussion in
\cite{Choi:2006qj,Choi:2009jt}) or as a $\U{1}_R$ explaining the hierarchy between the
Planck and the electroweak scales~\cite{Kappl:2008ie}.

The purpose of this review is to clarify the origin of discrete symmetries.
They can be obtained from continuous symmetries by spontaneous breaking. 
But this is not the only possibility. In fact, here we mainly focus on alternative
possibilities for the origin of discrete symmetries.  In section
\ref{sec:GeometricalOrigin} we discuss, in the framework of field theory, how
discrete symmetries can be related to the geometry of extra dimensions. The
discussion of higher--dimensional quantum field theories leaves certain
questions unanswered. We therefore change gear and present a top--down
derivation of discrete symmetries in section~\ref{sec:HeteroticOrbifolds},
focusing mainly on heterotic orbifolds, as they provide us with explicit
candidate models for a UV completion of the standard model, and, at the same
time, allow for a CFT description and hence for a detailed understanding of the
symmetries. As we shall see, the top--down settings are more restrictive than
the bottom--up models. Some of the restrictions can be thought of as
originating from the requirement of anomaly freedom, which we discuss
separately in section~\ref{sec:AnomalyFreedom}. Finally, we summarize our
discussion in section~\ref{sec:Summary}.

\section{Geometrical origin of discrete symmetries}
\label{sec:GeometricalOrigin}

In this section we present three possible origins of discrete symmetries. 
After briefly summarizing the standard approach and its limitations in 
section~\ref{sec:ZNfromU1}, we discuss how to obtain a discrete 
symmetry from extra dimensions, either as the symmetry of compact space 
(section~\ref{sec:GeometryFamily}) or as a remnant of higher dimensional 
Lorentz symmetry (section~\ref{sec:GeometryR}).

\subsection{Gauged discrete symmetries from continuous symmetries} 
\label{sec:ZNfromU1}

The perhaps most straightforward possibility for obtaining a discrete symmetry
is by spontaneous breaking of a continuous gauge symmetry.
As a simple example, consider a $\U{1}$
gauge group broken by the VEV of a scalar $\varphi$ with charge $q=3$.
Here we normalize the \U1 such that the charges are integer and have no
common divisor. The unbroken symmetry is given by those $\U{1}$
transformations that leave the vacuum invariant, i.e.\
\begin{equation}
 \mathrm{e}^{\I\, \alpha(x)\, q}\, \langle \varphi \rangle 
 ~=~ 
 \mathrm{e}^{3\I\, \alpha(x)}\, \langle \varphi \rangle 
 ~\stackrel{!}{=}~ 
 \langle \varphi \rangle 
 \quad\curvearrowright\quad 
 \alpha(x)~=~\frac{2\pi\, n}{3}\;,
\end{equation}
with $n = 0,1,2$. Hence, the (local) $\U{1}$ is broken to a (local) $\Z{3}$
subgroup. The extension of this discussion to the case of multiple \U1
factors which get broken by several VEVs is given in \cite{Petersen:2009ip}.

One may also get non--Abelian discrete symmetries by spontaneous breaking
(cf.\ e.g.\ \cite{Adulpravitchai:2009kd,Luhn:2011ip,Merle:2011vy}). However,
this typically involves very large representations of the corresponding
continuous symmetry, which often give rise to unwanted states in the broken
phase. Therefore, arguably, this possibility appears not to be too attractive. 
In what follows, we therefore discuss alternative possibilities in which the
discrete symmetries are related to the geometry of compact dimensions.
As we shall see, this scheme does not suffer from the above
problems, and is realized in explicit string--derived models of particle
physics.

\subsection{Repetition of families and symmetries}
\label{sec:GeometryFamily}

Discrete family symmetries can be motivated in settings with extra compact
dimensions. It is not surprising that such models offer an explanation for the
appearance of non--Abelian discrete flavor symmetries, because the latter are
symmetries of certain geometrical solids, which describe the compact 
dimensions. The symmetries of internal space
govern the interactions between fields that are localized in the compact 
dimensions and may eventually become flavor symmetries. 

The purpose of this subsection is to explain that (non--Abelian) family
symmetries can, to some extent, be understood geometrically.
Let us start with a very simple example with one extra compact dimension, 
the orbifold $\mathbbm{S}^1/\mathbbm{Z}_2$ (figure~\ref{fig:S1overZ2}). See 
appendix~\ref{sec:appendixA1} for a brief introduction to the construction 
of orbifolds.
\begin{figure}[h]
\centerline{\includegraphics{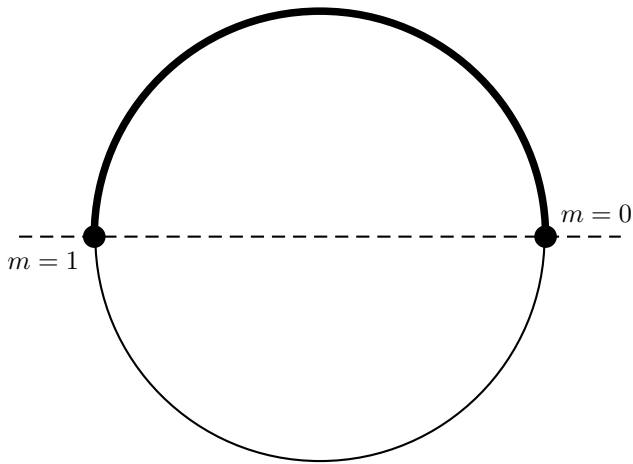}}
\caption{Example for one extra compact dimension: $\mathbbm{S}^1/\mathbbm{Z}_2$ 
orbifold. Points which are related by a reflection on the dashed line are 
identified. The fundamental region of the orbifold is an interval with the 
fixed points sitting at the boundaries.}
\label{fig:S1overZ2}
\end{figure}
This orbifold possesses two geometrically equivalent fixed points. Suppose
there are two states, i.e.\ two families of quarks and/or leptons,
$\psi_{m=0}$ and $\psi_{m=1}$, with identical quantum numbers, one of
them localized at each of the fixed points. Since the fixed points and the states 
$\psi_{m}$ are geometrically indistinguishable, there is an $S_2$ permutation 
symmetry relating them, which manifests itself as a symmetry of the theory.

\begin{figure}[h]
\centerline{\subfigure[$\mathbbm{T}^2_{\mathrm{SU}(3)}/\mathbbm{Z}_2\,.$]{%
    \includegraphics{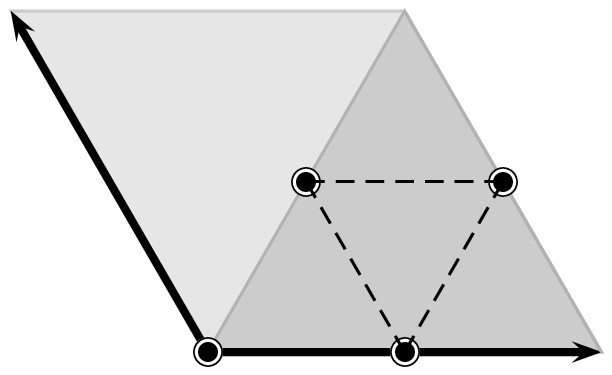}}
\qquad
\subfigure[Tetrahedron.]{\includegraphics[scale=0.7]{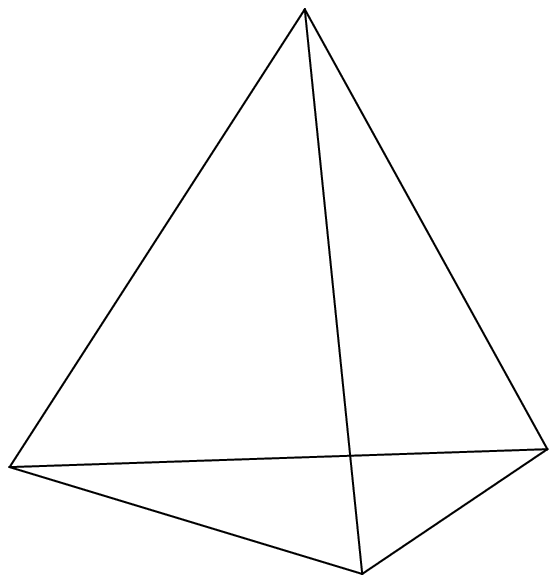}}}
\caption{Example for two extra compact dimensions: If the $\mathbbm{T}^2$ lattice vectors have equal length and enclose
$120^\circ$, one can also fold the fundamental region of
$\mathbbm{T}^2/\mathbbm{Z}_2$ (dark gray region in (a)) to the tetrahedron
 (b).}
\label{fig:tetrahedron}
\end{figure}

A somewhat more complex example is the tetrahedron in two extra compact 
dimensions (cf.\ \cite{Altarelli:2006kg}), which can be obtained from the
$\mathbbm{T}^2_{\mathrm{SU}(3)}/\mathbbm{Z}_2$ orbifold
(figure~\ref{fig:tetrahedron}). Here the subscript $\SU3$ indicates that the
basic translations defining the $\mathbbm{T}^2$ torus enjoy the same relations
as the simple roots of the Lie algebra of \SU3, i.e.\ enclose $120^\circ$ and
have equal lengths. 

Clearly, the tetrahedron is invariant under a discrete rotation by $120^\circ$
about an axis that goes through one corner and hits the opposite surface
orthogonally. There are four operations of this type represented by 
\begin{eqnarray}
T~= ~\left( \begin{array}{cccc} 1 & 0 & 0 & 0 \\
0 & 0 & 1 & 0 \\
0 & 0 & 0 & 1 \\
0 & 1 & 0 & 0
 \end{array}
\right)\;,\quad T\,S\;,\quad T\,S'\;,\quad T\,S\,S'\;,
\end{eqnarray}
in the basis where each of the four corners is represented by a 
four--dimensional vector $e_i$ with $(e_i)_j = \delta_{ij}$ and $i,j=1,2,3,4$. 
Furthermore,
\begin{subequations}
\label{eq:SandSprime}
\begin{eqnarray}
 S & = & \Id_{2 \times 2} \otimes \sigma_1 \;,\\
 S'& = & \sigma_1 \otimes \Id_{2 \times 2}
\end{eqnarray}
\end{subequations}
with the standard Pauli matrix $\sigma_1$. $T$ generates a  $\mathbbm{Z}_3$ and
$S$ generates a $\mathbbm{Z}_2$. In addition, one may allow for 
orientation--changing operations (with $\text{det}=-1$), for example, generated 
by $S''=\text{diag}(\Id_{2 \times 2}, \sigma_1)$.

Since these generators do not commute, the multiplicative closure yields a
non--Abelian discrete symmetry, being $S_4$. As mentioned in
\cite{Altarelli:2006kg}, if one restricts the allowed operations  to be
contained in proper Lorentz transformations, one arrives at the non--Abelian 
flavor symmetry generated by  $T$, $S$ and $S'$, which is $A_4$.  We therefore
arrive at the premature conclusion that, in a model in which each fixed point
carries a state, the family symmetry will be $A_4$. However, as pointed out in
\cite{Kobayashi:2006wq} and as we shall see later in more detail the actual
symmetry in UV complete settings is larger than that. 

In summary, we see that extra dimensions offer a compelling
explanation of non--Abelian discrete flavor symmetries. However, as the
settings discussed here are based on gauge theories in more than four
dimensions, one has to address the question of how to complete them in the UV.
We will come back to this question in section~\ref{sec:HeteroticOrbifolds},
where we will see that string models indeed often exhibit non--Abelian
discrete family symmetries.

\subsection{Discrete $\boldsymbol{R}$ symmetries}
\label{sec:GeometryR}

In supersymmetric theories there are the so--called $R$ symmetries which, by
definition, do not commute with supersymmetries. Such symmetries can originate
from extra dimensions as well. Specifically they are (discrete) remnants of the
Lorentz symmetry of compact dimensions. The perhaps simplest way of seeing this
is by recalling that under Lorentz rotations spinors, vectors and scalars
transform differently such that different parts of superfields have different
charges. This means, in particular, that $R$ symmetries are deeply connected to
the fundamental symmetries of space--time. 

Let us illustrate this point in more detail by discussing toy--settings with 
two compact dimensions (without discussing SUSY breaking). If these dimensions were 
flat (and infinite) the setup would exhibit an SO(2) rotation symmetry.  For 
instance, this symmetry can be defined by its action on the extra components of 
the gauge fields,
\begin{equation}\label{eq:Rotation1}
 \left(\begin{array}{c}A_5\\ A_6\end{array}\right)
 ~\to~
 \left(\begin{array}{cc}
  \cos\zeta & -\sin\zeta\\
  \sin\zeta & \cos\zeta
 \end{array}\right)\,
 \left(\begin{array}{c}A_5\\ A_6\end{array}\right)\;.
\end{equation}
Since such components get combined to the scalar component of a chiral
superfield, describing a bulk field (or an untwisted sector field in 
string--derived orbifolds), it is more convenient to recast
\eqref{eq:Rotation1} in complex notation,
\begin{equation}\label{eq:U156extravector}
 \mathrm{U}(1)_{56}~:~
 A_5+\I A_6~\to~\mathrm{e}^{\I\,\zeta}\,(A_5+\I A_6)\;.
\end{equation}
On the other hand, the spinor component of this `untwisted superfield' turns
out to transform differently under the Lorentz group. To understand this, note
that the 4D spinor $\rho$ is contained in the higher--dimensional one ($\Psi$)
according to $\Psi~=~\rho\otimes\chi$,
where $\chi$ is a spinorial zero mode in internal space. Recalling that spinors
always rotate half as quickly as vectors under Lorentz transformations leads to
the transformation law
\begin{equation}\label{eq:U156spinor}
  \mathrm{U}(1)_{56}~:~
 \rho~\to~\mathrm{e}^{\I\,\zeta/2}\,\rho\;.
\end{equation}
In the 4D superfield 
\begin{equation}
 \Phi~=~\frac{1}{\sqrt{2}}\,(A_5+\I\,A_6)+\sqrt{2}\,\theta\rho+
 \theta\theta\,F
\end{equation}
the superspace coordinates $\theta$ balance the transformations of the
components \eqref{eq:U156extravector} and \eqref{eq:U156spinor}, i.e.\
\begin{equation}
 \mathrm{U}(1)_{56}\::\:
 \theta ~\to~
 \mathrm{e}^{\I\,\zeta/2}\,\theta\;.
\end{equation}
Hence, $\mathrm{U}(1)_{56}$ originating from the 6D Lorentz symmetry denotes 
an $R$ symmetry.

It is also clear that typically a compact space does not possess the full 
Lorentz symmetry. For example, orbifolds can have discrete rotational symmetries 
and hence can naturally provide discrete $R$ symmetries, see section~\ref{sec:StringR} 
for more details in the case of string compactifications on orbifolds.

\section{Orbifolds and string selection rules}
\label{sec:HeteroticOrbifolds}

\begin{figure}[h]
 \centerline{\includegraphics{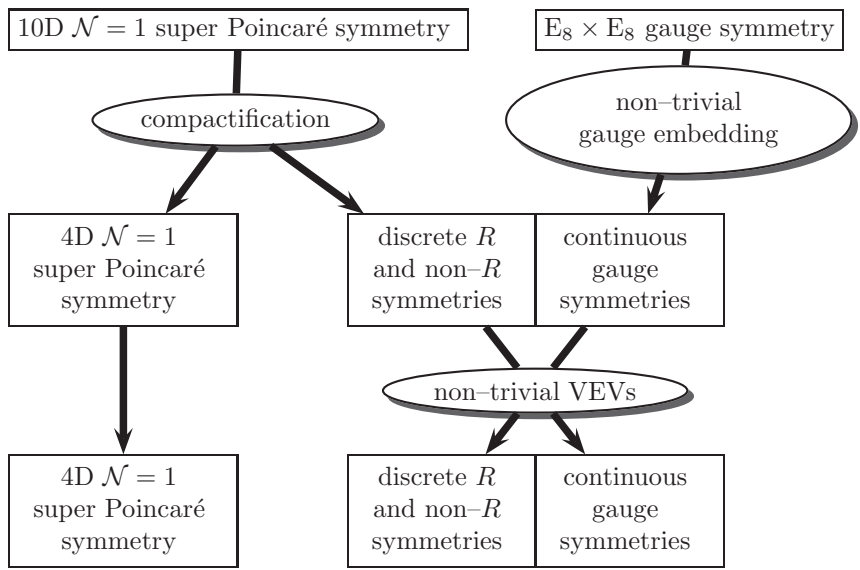}}
  \caption{Origin of symmetries in heterotic orbifold compactifications.
  By compactification of six dimensions and appropriate gauge
  embedding the 10D super Poincar\'e and $\E8\times\E8$ symmetries get broken
  to the 4D super Poincar\'e, a 4D gauge and various discrete $R$ and non--$R$
  symmetries. The latter two get further broken to subgroups by non--trivial
  VEVs of certain charged fields.
 }
\label{fig:origin}
\end{figure}

So far, our discussion was purely bottom--up. It is, however, instructive to
comment on the situation in top--down models. The geometrical repetition of
families, as briefly discussed in section~\ref{sec:GeometryFamily}, is a common
feature of most string compactifications.
\begin{enumerate}
 \item In heterotic orbifolds, very often families come from so--called
 twisted sectors, which correspond to states localized
 at the orbifold fixed points in the extra dimensions.
 We will discuss the emergent family symmetries in more detail below.
 \item In $D$--brane models (see e.g.\ \cite{Blumenhagen:2006ci} for a review)
 the repetition of families is due to the fact that branes can wrap cycles 
 (i.e.\ some directions in the extra dimensions)  multiple times. Therefore,
 one can have non--trivial intersection numbers between different branes,
 leading to otherwise equivalent chiral states localized at the
 intersections. Therefore such models also generically exhibit non--trivial
 family symmetries.
 Also $F$ theory models have non--trivial family symmetries, which often
 lead to the problem that the Yukawa couplings have rank one \cite{Heckman:2008qa}.
\end{enumerate}
In what follows, we will focus on the heterotic string compactified on
(toroidal) orbifolds. There are two main reasons for this choice. First of all,
the heterotic framework gives rise to explicit globally consistent candidate
models for physics beyond the standard model
\cite{Lebedev:2006kn,Lebedev:2008un,Anderson:2011ns, Anderson:2012yf}. Second, at the same time, this scheme is
simple enough to fully understand the symmetries. Discrete symmetries can
appear mainly in two ways: (i) from the compacification to 4D as remnants of
higher dimensional gauge/Lorentz symmetry and (ii) from going to a special
vacuum configuration where some of the fields of the 4D effective theory obtain
VEVs and hence induce further symmetry breaking. The situation is schematically
illustrated in figure~\ref{fig:origin}. 

Orbifolds are six--dimensional compact spaces which, in contrast to a general
Calabi--Yau compactification, have additional discrete symmetries which manifest
themselves in the four-dimensional effective theory. Brief introductions 
to heterotic orbifolds and to the selection rules that govern the allowed 
terms of the superpotential are given in appendix~\ref{sec:appendixA}.

\subsection{Discrete symmetries from string selection rules}

\subsubsection{Abelian symmetries}
\label{sec:abeliansymmetries}

In general, there are two possible origins for Abelian (non--$R$) discrete
symmetries in heterotic orbifold compactifications. Either they can arise from
the space group selection rule discussed in
appendix~\ref{sec:stringselectionrules} or as a discrete remnant of a
spontaneously broken gauge symmetry. The second possibility was discussed in 
section~\ref{sec:ZNfromU1}, the first one will be presented in the following.

For the sake of concreteness, we consider the
$\mathbbm{S}^1/\Z{2}$ orbicircle of section~\ref{sec:GeometricalOrigin}. In
this case, the space group consists of the elements $\left(\theta^k, m\,
e\right)$, where $\theta=-1$ and $k\in\{0,1\}$ describe the $\Z{2}$ reflection, $m\in\Z{}$, $e=2\pi\, R$ and $R$
denotes the radius of the circle $\mathbbm{S}^1$. As
illustrated in figure~\ref{fig:S1overZ2}, the integer $m$ specifies the
location of the twisted state (or `brane field'), which have $k=1$, as opposed
to untwisted states (or `bulk fields') which have $k=0$. The space group
selection rule requires the product of space--group elements of the states
involved in a coupling to be congruent to identity (see
appendix~\ref{sec:stringselectionrules}). This gives rise to an Abelian
$\Z{2}^k\times\Z{2}^m$ symmetry, i.e.\ 
\begin{equation}
\label{eqn:AbelianPartOrbicircle}
 \prod_{r=1}^L g_r~=~ \Id \quad\curvearrowright\quad 
 \left\{\begin{array}{lcl}
 \Z2^k & : & \displaystyle\sum_{r=1}^L k^{(r)}~=~0\mod 2\;,\\
 \Z2^m & : & \displaystyle\sum_{r=1}^L m^{(r)}~=~0\mod 2\;.
 \end{array}\right.
\end{equation}
We will refer to the condition on $k^{(r)}$ as the point group selection rule
and to the second one on $m^{(r)}$ as the $m$--rule.

\subsubsection{Non--Abelian symmetries}
\label{sec:nonabeliansymmetries}

A particularly interesting situation arises if the two fixed points at $m=0$ and
$1$ are equivalent, which happens to be the case  unless one introduces a
non--trivial background field (either a so--called  discrete Wilson line
\cite{Ibanez:1986tp} or the $B$-field (discrete torsion) \cite{Vafa:1986wx}). 
In this case there is an additional $S_2$ permutation symmetry that 
interchanges $m=0$ and $m=1$. As we shall discuss now, together with the
$\Z{2}^k\times\Z{2}^m$ symmetry discussed above in
section~\ref{sec:abeliansymmetries}, this leads to a non--Abelian discrete
symmetry $D_4$ \cite{Dixon:1986qv,Kobayashi:2006wq}.

We combine a state from the fixed point at $m=0$ and a state from the one
at $m=1$ into a two--dimensional vector, i.e.\ a doublet. From
equation~\ref{eqn:AbelianPartOrbicircle} we see that the space group
selection rule is generated in this basis by the elements
\begin{equation}
 -\Id_{2\times2}~=~\left(\begin{array}{cc}-1 & 0 \\ 0 & -1\\\end{array}\right)\;, \quad
 \sigma_3~ =~ \left(\begin{array}{cc}1 & 0 \\ 0 & -1\\\end{array}\right)\;,
\end{equation}
i.e.\ the element $-\Id_{2 \times 2}$ generates $\Z{2}^k$, i.e.\ the
point group selection rule, and $\sigma_3$ generates $\Z{2}^m$, i.e.\ the
$m$--rule. The additional element that generates the permutation of the two
states is given by
\begin{equation}
 \sigma_1~=~\left(\begin{array}{cc}0 & 1 \\ 1 & 0\\\end{array}\right)\;.
\end{equation}
The multiplicative closure of these three elements yields a non--Abelian group with eight
elements $\{\Id_{2 \times 2}, -\Id_{2 \times 2}, \pm\sigma_1,
\pm\sigma_3, \pm \I \sigma_2\}$ and is known as the dihedral group $D_4$,
associated with the symmetry of a square.

Similar to the $\mathbbm{S}^1/\Z{2}$ case, the $\mathbbm{T}^2/\Z{2}$ orbifold
without Wilson lines (see figure~\ref{fig:fixedpoints}) generically has a
$(D_4\times D_4)/\Z{2}$ flavor symmetry which originates from the Abelian space
group selection rule $\Z{2}^3$ combined with the permutation symmetries $S$ and
$S'$ of equation~\ref{eq:SandSprime}. It can be enhanced further for
special values of the angle and the two radii of $\mathbbm{T}^2/\Z{2}$. For
example, when the orbifold geometrically is a tetrahedron the naive geometrical 
$S_4$ symmetry obtained from field theory considerations in section~\ref{sec:GeometryFamily} 
gets enhanced to $\text{SW}_4$, which has 192 elements, by the stringy space group selection rule 
\cite{Kobayashi:2006wq}.  
If one allows only for proper Lorentz transformations, one obtains a group with 
96 elements which is contained in $\text{SW}_4$. The string description allows 
us to clarify whether or not one should consider operations which are not 
contained in the proper Lorentz transformations. The couplings between states 
localized at different fixed points go like $\mathrm{e}^{-a\,T}$, where $T$ 
denotes the K\"ahler modulus of the corresponding orbifold plane. The real part 
of $T$ is proportional to $R^2$, where $R$ is the radius of the underlying 
torus. Clearly, $R^2$ does not change under these extra reflections, such that 
the absolute values of the coupling strengths will enjoy the larger symmetry 
$\text{SW}_4$. On the other hand, the imaginary part of $T$, the so--called 
$T$--axion, is related to the anti--symmetric tensor field in compact space, 
and does change its sign under the extra reflections. Hence, if the $T$--axion 
acquires a non--trivial VEV, the phases of the coupling strengths do no longer 
enjoy the larger symmetries. As is well known, breaking the reflection 
symmetries in internal space can be related to CP violation in the effective 4D 
theory (cf.\ \cite{Dine:1992ya,Kobayashi:2003gf}), and what we discussed here 
is just an example for this statement. Note that there are different 
possibilities to obtain non--trivial CP phases, also based on non--Abelian 
discrete symmetries (cf.\ \cite{Chen:2009gf}). It should be interesting to see 
if these also have an interpretation in terms of reflection symmetries in 
compact space.

 Different lower--dimensional building blocks of orbifolds lead to other
non--Abelian discrete symmetries (table~\ref{tab:Survey}). 
\begin{table}[h]
\begin{center}
\begin{tabular}{|c|c|c|c|}  \hline
orbifold & flavor symmetry & sector &  string fundamental states \\ \hline
$\mathbbm{S}^1/\mathbbm{Z}_2$ & $D_4$  
& $U$ & $\boldsymbol{1}$ \\
   &  & $T_1$ & {\bf  2} \\
\hline $\mathbbm{T}^2/\mathbbm{Z}_2$ & $(D_4\times D_4)/\mathbbm{Z}_2$    
       & $U$   & $\boldsymbol{1}$ \\
   &   & $T_1$ & $\boldsymbol{4}$ \\  \hline
$\mathbbm{T}^2/\mathbbm{Z}_3$ & $\Delta(54)$   
       & $U$   & $\boldsymbol{1}$ \\
&      & $T_1$ & ${\bf 3}$ \\
&      & $T_2$ & ${\bar {\bf 3}}$ \\ \hline
$\mathbbm{T}^2/\mathbbm{Z}_4$ &   & $U$ & $\boldsymbol{1}$ \\
   & $(D_4 \times \mathbbm{Z}_4)/\mathbbm{Z}_2$  & $T_1$ & {\bf 2} \\
   & & $T_2$ & $\boldsymbol{1}_{A_1} + \boldsymbol{1}_{B_1} +\boldsymbol{1}_{B_2} + \boldsymbol{1}_{A_2}$ \\
   & & $T_3$ & {\bf 2} \\
\hline
$\mathbbm{T}^4/\mathbbm{Z}_8$ &  & $U$ &
$\boldsymbol{1}$ \\
   &    & $T_1$ & {\bf 2} \\
   &  $(D_4 \times \mathbbm{Z}_8) / \mathbbm{Z}_2$  & $T_2$ &
$ \boldsymbol{1}_{A_1} + \boldsymbol{1}_{B_1} +\boldsymbol{1}_{B_2} + \boldsymbol{1}_{A_2}$ \\
   & & $T_3$ & {\bf 2} \\
   & & $T_4$ &
$4 \times (\boldsymbol{1}_{A_1} + \boldsymbol{1}_{B_1}
+\boldsymbol{1}_{B_2} + \boldsymbol{1}_{A_2})$
\\  \hline
$\mathbbm{T}^4/\mathbbm{Z}_{12}$ & trivial & & \\ \hline
$\mathbbm{T}^6/\mathbbm{Z}_7$ & & $U$ & $\boldsymbol{1}$ \\
    & $S_7 \ltimes (\mathbbm{Z}_7)^6$ & $T_k$ &
    $\boldsymbol{7}$ \\
   & & $T_{7-k}$ & ${\bar {\bf 7}}$ \\
\hline
\end{tabular}
\end{center}
\caption{Survey of flavor symmetries arising from building blocks of orbifolds
(from \cite{Kobayashi:2006wq}). The $T_k$ denote the various twisted sectors 
 and $U$ the untwisted sector.}
\label{tab:Survey}
\end{table}

The (non--Abelian) flavor symmetry could be broken in two ways: (i)
explicitly: the presence of orbifold Wilson lines breaks the
permutation symmetry, at least partially. If the permutation symmetry is
completely broken, the remaining flavor group is Abelian. (ii)
spontaneously: by the VEV of some twisted field, since twisted
fields necessarily transform in a non--singlet representation under the flavor
group.  For example, $\Delta(54)$ can be broken to $S_3$ by the VEV of a
triplet  $\boldsymbol{3}$ (e.g.\ $\langle\boldsymbol{3}\rangle=(v,v,v)$).

\subsubsection{$\boldsymbol{R}$ symmetries}
\label{sec:StringR}

As already mentioned, discrete $R$ symmetries could arise as discrete remnants
of the Lorentz symmetry of compact dimensions.  This is also true for
string--derived orbifold models.

What are the $R$ charges of states localized at the fixed points? In the
framework of field theory one cannot answer this question unambiguously. For
instance, in many field--theoretic analysis the profiles of these fields are
taken to be $\delta$--functions with support at the fixed points, from which one
may conclude that the states transform trivially under the discrete $R$
symmetries. It turns out that the naive field--theoretic expectation is
incorrect. However, in string theory one can address this question. 
Specifically, in heterotic orbifolds the $R$ symmetries derive from the
$H$--momentum conservation law and one can determine the $R$ charges
unambiguously. We will discuss an explicit example in
section~\ref{sec:NonPerturbativeViolation}.

\subsection{Discrete symmetries in explicit models}

Having seen how discrete symmetries arise in the effective field--theoretic
description, we will now discuss which symmetries appear in explicit
string models. 

\subsubsection{Flavor symmetries} 

In recent years, many MSSM candidate models have emerged from heterotic
orbifolds 
\cite{Buchmuller:2005jr,Buchmuller:2006ik,Lebedev:2006kn,Lebedev:2007hv}, known
as the ``heterotic mini--landscape''. These models have a common flavor 
structure: focusing on the two--torus where a $\Z{2}$ acts, the two light 
generations are localized on equivalent fixed points and the third one is in the 
bulk.
Therefore, as discussed above, there is a $D_4$ flavor symmetry, under which the
two light generations transform as a doublet whereas the the third family
transforms trivially (Let us mention that there are also alternative models
without this $D_4$ \cite{Kim:2007mt, Lebedev:2008un}).  This symmetry is broken
in potentially realistic vacua by the VEVs of some localized  singlets. Yet,
using the $D_4$ symmetric situation as a starting point and then considering 
corrections can have certain advantages when discussing the (supersymmetric)
flavor structure (cf.\ \cite{Ko:2007dz}). The emerging scheme is somewhat similar
to the one of `minimal flavor violation'
\cite{Chivukula:1987py,Buras:2000dm,D'Ambrosio:2002ex}. In particular, the
structure of the soft masses is
\begin{equation}
 \widetilde{m}^2~=~\left(\begin{array}{ccc}
  a &  0 & 0\\
  0 &  a & 0\\ 
  0 &  0 & b
 \end{array}\right)+\text{terms proportional to $D_4$ breaking VEVs}\;.
\end{equation}
It is known that such an approximate form of the soft masses makes it possible
to avoid the supersymmetric flavor problems. In addition, it naturally 
allows for scenarios in which the third family of squarks and sleptons is
substantially lighter than the first two generations of superpartners (cf.\ the
discussion in \cite{Krippendorf:2012ir}).

\subsubsection{Flavor--independent symmetries}

In grand unified models, matter or $R$ parity can be obtained 
from baryon--minus--lepton--number symmetry $\U1_{B-L}$ by spontaneous
breaking, and the same is true in string--derived models \cite{Lebedev:2007hv},
the only difference being that $\U1_{B-L}$ is not in GUT normalization and no
large representations (such as $\overline{\boldsymbol{126}}$--plets of \SO{10})
are required (nor available) to achieve the breaking
$\U1_{B-L}\to\Z2^\mathcal{M}$.
That is, string theory avoids huge representations like the
$\overline{\boldsymbol{126}}$--plets, but still allows us to derive matter
parity from a local $B-L$ symmetry. 

Similarly, proton hexality can be obtained from Pati--Salam (PS) times an 
extra $\U{1}$ symmetry \cite{Forste:2010pf}. Explicit orbifold models from 
$\Z{4}\times\Z{4}$ compactifications using a local GUT approach, 
\begin{equation}
\E{8}\text{ in 10D}\to\SO{12}\text{ in 6D}\to\text{PS}\times\U{1}\to\text{SM in 4D}\;,
\end{equation}
revealed 850 heterotic MSSMs (i.e.\ three generations of quarks and leptons plus 
vector--like exotics), many of them with the correct proton hexality charge 
assignment for at least some quarks and leptons~\cite{Forste:2010pf}.

\subsubsection{$\boldsymbol{R}$ symmetries} 
\label{sec:ExplicitModelsR}

$R$ symmetries play an important role in string models. In
particular, approximate continuous $R$ symmetries, which derive from exact
discrete $R$ symmetries, can explain the large hierarchy between the Planck,
GUT and/or string scales on the one hand and the electroweak and/or
supersymmetry breaking scales on the other hand. It has
been demonstrated that, in the presence of a continuous $R$ symmetry, at field
configurations that satisfy the $F$--term constraints, the VEV of the superpotential
vanishes \cite{Kappl:2008ie}. If there is an approximate $R$ symmetry that gets
explicitly broken at some high order $N$, the vacuum expectation value of the
superpotential, or equivalently the gravitino mass $m_{3/2}$, goes like
\begin{equation}\label{eq:WKappl}
\langle\mathscr{W}\rangle~\sim~\langle s\rangle^N\;,
\end{equation}
where $\langle s\rangle$ denotes a typical size of a VEV of fields that break
the symmetry spontaneously (in Planck units) and $N$ is of the order 10 in
explicit  examples. Further, in the context of the MSSM it has been shown that
in settings in which matter charges are consistent with grand unification, the
only anomaly--free symmetries that can forbid the $\mu$ term are $R$ symmetries
\cite{Lee:2011dya}. Given that $\langle\mathscr{W}\rangle$, or, equivalently
$m_{3/2}$, is the order parameter of $R$ symmetry breaking, this yields a
relation between $\mu$ and $m_{3/2}$ \cite{Kappl:2008ie,Brummer:2010fr}, i.e.\
constitutes a solution to the $\mu$ problem. This solution does, unlike the
Giudice--Masiero mechanism \cite{Giudice:1988yz}, not rely on a specific
structure of the K\"ahler potential, rather it provides a holomorphic $\mu$ term
of the right size, similar to the Kim--Nilles picture \cite{Kim:1983dt}.

\section{Anomaly Freedom}
\label{sec:AnomalyFreedom}

\subsection{Anomaly constraints vs.\ embedding constraints}

How can one derive anomaly constraints on discrete symmetries? It is instructive
to review how they have been derived in the past. Ib\'a\~nez and Ross
\cite{Ibanez:1991hv} have used the following strategy: they have obtained
$\Z{N}$ symmetries from \U1 by spontaneous breaking, as discussed in
section~\ref{sec:ZNfromU1}. It is obvious that, if the \U1 is non--anomalous,
and the spontaneous breaking is done consistently, then also \Z{N} is
anomaly--free. However, one may question whether these are in general true anomaly
constraints or rather embedding constraints, i.e.\ constraints that restrict  
the choice of the non--anomalous continuous gauge group into which the discrete group is 
supposed to be embedded. 

Araki \cite{Araki:2006mw} proposed an alternative derivation of the anomaly 
constraints, which does not rely on embedding the discrete symmetry into a 
continuous one, but by using the path integral method \cite{Fujikawa:1979ay}. This 
strategy has been applied to the \Z{N} case \cite{Araki:2008ek} with the result 
that all Ib\'a\~nez--Ross constraints apply except for the $\Z{N}^3$ ones, 
which are known not to constitute true anomaly constraints 
\cite{Banks:1991xj,Lee:2011dya}.

Also discrete anomaly constraints for non--Abelian discrete symmetries 
have first been derived by using the embedding strategy \cite{Frampton:1994rk} 
(see \cite{Luhn:2008sa} for a more recent discussion).
While, again, these constraints ensure anomaly freedom, they turn
out to be, in general, not true anomaly constraints but rather embedding
constraints. That is, if these constraints are satisfied, the symmetry is anomaly 
free, but the converse is not necessarily true. In particular, the constraints can 
depend on the choice of the continuous symmetry into which the discrete one is 
supposed to be embedded. The true constraints can be derived with the path integral 
method \cite{Araki:2006mw}, and one finds that one only has to check anomaly freedom
for the Abelian subgroups of a given non--Abelian symmetry
\cite{Araki:2006mw,Araki:2008ek}. For a discrete group $D$ and a continuous
gauge symmetry $G$ one obtains the conditions that
\begin{eqnarray}
 \sum_{(\boldsymbol{r}^{(f)},\boldsymbol{d}^{(f)})} 
 \delta^{(f)}\cdot \ell(\boldsymbol{r}^{(f)})
 & \stackrel{!}{=} &
 0\mod \frac{N}{2}\;,\label{eq:NAcondition-gauge2}
\end{eqnarray}
where the sum `$\sum_{(\boldsymbol{r}^{(f)},\boldsymbol{d}^{(f)})} $'
is over representations which are non--trivial w.r.t.\
to both $G$ and $D$.
The discrete Abelian charge, denoted by $\delta^{(f)}$,
can be expressed in terms of the group elements $U(\boldsymbol{d}^{(f)})$ as 
\begin{equation}\label{eq:trtaui}
 \delta^{(f)}~=~N\,\frac{\ln \det U(\boldsymbol{d}^{(f)})}{2\pi\,\I}\;.
\end{equation}
For the mixed gravitational--$D$ anomaly one finds
\begin{eqnarray}
 \sum_{\boldsymbol{d}^{(f)}}\delta^{(f)}
 & \stackrel{!}{=} &
 0\mod \frac{N}{2}\;,
 \label{eq:NAcondition-grav2}
\end{eqnarray}
where the symbol `$\sum_{\boldsymbol{d}^{(f)}}$' means that the sum extends over all non--trivial
representations $\boldsymbol{d}^{(f)}$ of $D$.
What does it mean if a given discrete symmetry does not satisfy
these constraints? In general, one may argue that in such a case
the symmetry will be broken in an uncontrollable way and all 
the predictive power of the (discrete) symmetry will be lost. 
For useful applications in particle physics, reliable discrete
symmetries should thus be anomaly free. There is, however, an
exception: for the anomalous symmetry the anomalies might be 
cancelled (microscopically) by a discrete Green-Schwarz mechanism. 
In what follows, we shall discuss this possibility in detail.

\subsection{Non--perturbative ``violation'' of discrete symmetries and discrete
Green--Schwarz anomaly cancellation}
\label{sec:NonPerturbativeViolation}

As in the case of continuous symmetries, discrete anomalies can be cancelled by
a Green--Schwarz (GS) mechanism (for a discussion in the path integral formalism
see \cite{Lee:2011dya}). Also here this requires the presence of a scalar, the
GS axion, which multiplies some $F_{\mu\nu}\widetilde{F}^{\mu\nu}$ terms (with
$F^{\mu\nu}$ denoting the field strength of some continuous gauge symmetry of
the model), and shifts under the discrete symmetry. Once the axion acquires its
vacuum expectation value, the discrete symmetry gets broken spontaneously.
Effectively this leads to a situation in which the (anomalous part of the)
discrete group appears to be broken by non--perturbative
effects.\footnote{Non--perturbative effects generate couplings of the form 
$\exp (-\I a) \phi_1 \ldots \phi_n$, where $a$ denotes the GS axion and the 
$\phi_i$ some (matter) fields of the theory. Such terms are invariant under 
the full discrete group when one takes the shift transformation of the 
GS axion $a$ into account. But, when $a$ obtains a vacuum expectation value, the 
(`anomalous' part of the) discrete group is broken spontaneously.}

As an example, consider the $\Z{4}^R$ symmetry discussed in
\cite{Lee:2010gv,Lee:2011dya}. It forbids the $\mu$ term and dimension 4 and 5
proton decay operators at the perturbative level. It appears to be broken by
non--perturbative effects to its `non--anomalous' subgroup, i.e.\ to
$\Z{2}^\mathcal{M}$ matter parity. The order parameter of this $R$ symmetry 
breaking is the vacuum expectation value of the superpotential, i.e.\ the
gravitino mass. One therefore has, in the context of gravity mediation, a $\mu$
term of the correct size (cf.\ the analogous discussion in 
section~\ref{sec:ExplicitModelsR}) while dimension five proton decay remains far below
the experimental limits.

Similar to the case of $R$ symmetries, also non--$R$ symmetries
can appear anomalous and hence be broken non--perturbatively. This, again, introduces a
hierarchically small breaking of the discrete symmetry. It remains to be seen
whether this mechanism can provide us with solutions to some of the open questions in
flavor physics.

\section{Summary}
\label{sec:Summary}

The flavor structure of the SM remains one of the greatest puzzles in particle
physics. Flavor symmetries appear to be instrumental for solving this puzzle.
Optimistically one may hope to find a compelling model that explains the
observed flavor structure. In this case the question where the underlying family
symmetries originate from is of greatest importance since given a deeper
understanding may allow us to relate the observed fermion masses and mixing to
some fundamental properties of our world.

In this paper we have reviewed the possible origin of discrete symmetries,
paying particular attention to discrete flavor symmetries. Discrete symmetries
can arise from continuous symmetries by spontaneous breaking or from extra
dimensions. While for Abelian symmetries the first option is a very common tool
in model building, we have argued that obtaining non--Abelian discrete
symmetries from continuous ones (in four dimensions) does not lead to compelling
models. On the other hand, non--Abelian discrete symmetries do arise in models
with extra dimensions, where they are deeply connected to the explanation of the
repetition of families. In particular, in stringy extensions of the standard
model such symmetries often arise. Therefore they can play an important role in
understanding or addressing the flavor puzzle in the standard model as well as
in solving flavor problems in extensions such as the MSSM. 

We have also commented on discrete anomalies, which constrain possible discrete
symmetries in bottom--up model building. As we have pointed out, one should
carefully distinguish between embedding constraints and true anomaly
constraints. Discrete symmetries that appear anomalous open very attractive
possibilities in model building as they appear to be broken non--perturbatively,
i.e.\ the breaking can be hierarchically small. This observation has been
applied to the $\mu$ parameter of the MSSM. It remains to be seen whether
hierarchies in flavor physics can have a similar explanation.

\subsection*{Acknowledgments}

One of us (M.R.) would like to thank Mu--Chun Chen for useful discussions and the 
UC Irvine, where part of this work was done, for hospitality.
This work was partially supported by the SFB--Transregio TR33 ``The Dark
Universe'' (Deutsche Forschungsgemeinschaft), the SFB 676, the European Union 7th network
program ``Unification in the LHC era'' (PITN--GA--2009--237920) and the DFG
cluster of excellence ``Origin and Structure of the Universe'' (Deutsche
Forschungsgemeinschaft).

\appendix

\section{Orbifolds}
\label{sec:appendixA}

We give a brief introduction to orbifolds following \cite{Vaudrevange:2008sm,
RamosSanchez:2008tn}. We start with the geometrical construction in
appendix~\ref{sec:appendixA1}. In appendix~\ref{sec:appendixA2} we
depict how heterotic strings are compactified on orbifolds and
appendix~\ref{sec:stringselectionrules} reviews string selection rules.

\subsection{Construction of orbifolds}
\label{sec:appendixA1}

\begin{figure}[h]
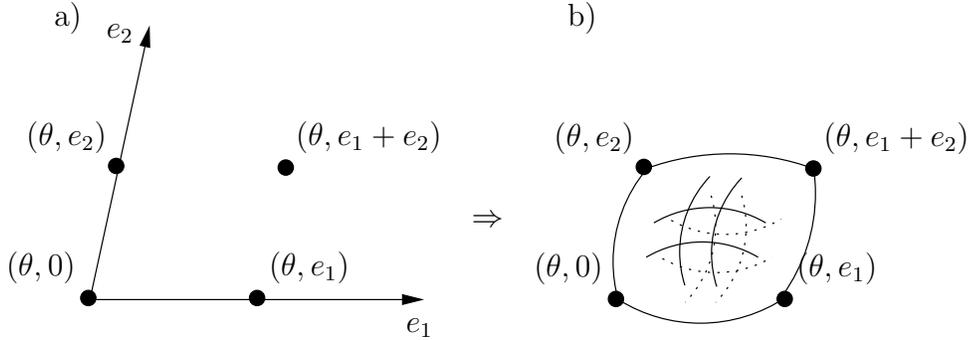

  \centerline{\input fixedpoints.pstex_t}
  \caption{a) The four fixed points labeled by their constructing elements for a
two--dimensional $\mathbbm{T}^2/\Z{2}$ example (i.e.\ with $\theta=-\Id_{2
\times 2}$). b) The orbifold of a) folded up to a pillow--like object with fixed
points at the corners of the pillow.}
\label{fig:fixedpoints}
\end{figure}

From the geometrical point of view, a $d$--dimensional (toroidal) orbifold is defined as the
quotient of $\mathbbm{R}^d$ divided by a discrete group $S$, called the space
group. For $\Z{N}$ orbifolds, the elements of the space group $g \in S$ are
given by
\begin{equation}
 g~=~\left(\theta^k, n_\alpha\, e_\alpha\right) 
 \qquad\text{and act as}\qquad 
 g\, X~=~\theta^k\, X + n_\alpha\, e_\alpha\;,
\end{equation}
with sum over $\alpha=1,\ldots,d$ and $X \in \mathbbm{R}^d$. The $d$  (linearly
independent) vectors $e_\alpha$ generate a lattice $\Gamma$ and hence define a
torus $\mathbbm{T}^d$. The rotation $\theta$ is of order $N$ (i.e.\ $\theta^N =
\Id$) and is chosen to be an automorphism of $\Gamma$. Then the action of $S$ is
not free, i.e.\ there are fixed points $X_g\in\mathbbm{R}^d$ with $g X_g = X_g$
for some $g \in S$. The space group element $g$ associated to the fixed point
$X_g$ is called the constructing element, see figure~\ref{fig:fixedpoints}. The
resulting orbifold is written as $\mathbbm{T}^d/\Z{N}$.

\subsection{Strings on orbifolds}
\label{sec:appendixA2}

Compactifying the heterotic string on six--dimensional orbifolds yields three
different classes of closed strings: (i) untwisted strings with constructing
element $\left(\Id, 0\right)$ which would also close in uncompactified
space, (ii) winding modes with constructing elements $\left(\Id, n_\alpha
e_\alpha\right)$ which would also close on the torus and (iii) twisted strings,
localized at the fixed points, with constructing elements $\left(\theta^k,
n_\alpha e_\alpha\right)$ with $k \neq 0$ which only close on the orbifold due
to the $\theta$ rotation. The winding modes are massive with masses near the
Planck scale. Since we are only interested in the low--energy effective action
they are ignored in the following. 

The geometrical action of the space group has to be amended by an action
on the gauge degrees of freedom of the heterotic string in order to fulfill the
stringy consistency conditions of modular invariance. In the standard approach
this is achieved by so called shifts and Wilson lines. Specifying these input
parameters completely defines an orbifold compactification and allows one to
compute the massless spectrum. An elegant way to obtain consistent orbifold
models, for example MSSM--like models, to compute their massless spectra and to
analyze their resulting four--dimensional effective theories is given by the
public code ``Orbifolder''~\cite{Nilles:2011aj}.

\subsection{String selection rules}
\label{sec:stringselectionrules}

The CFT description allows one to compute scattering amplitudes of
strings on orbifolds. In the four--dimensional effective theory these amplitudes
enter as coupling strengths of allowed terms in the superpotential. Their
computation is technically involved. Hence, at a first step one is only
interested in the string selection rules determining which coupling is allowed
or forbidden. In many cases the string selection rules can be interpreted as a
symmetry of the four--dimensional effective theory. The (standard) string selection rules are:

\begin{enumerate}
\item{Gauge invariance}
\item{Space group selection rule: } The space group selection rule reflects the geometrical possibility of orbifold strings to join. Consider $L$ strings with constructing elements $g_r = \left(\theta^{k^{(r)}}, n_\alpha^{(r)}e_\alpha\right)$. Then the coupling is allowed if $\prod_{r=1}^L g_r = \Id$, see
figure~\ref{fig:spacegroupselectionrule}.

\begin{figure}[t]
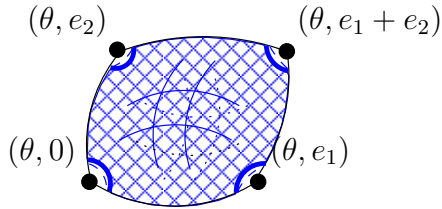

  \centerline{\input spacegroupselectionrule.pstex_t}
  \caption{Visualization of the space group selection rule for four twisted
   states, indicated by the four bold strings at the four fixed points, see
   figure~\ref{fig:fixedpoints}. The condition $\left(\theta,0\right)
   \left(\theta,e_1\right) \left(\theta,e_2\right) \left(\theta,e_1+e_2\right) =
   \left(\Id, (\Id-\theta)\Gamma\right)$ is satisfied and hence the coupling is
   allowed.}
\label{fig:spacegroupselectionrule}
\end{figure}

\item{$R$ charge conservation: }
$R$ charge conservation is a discrete remnant of ten--dimensional Lorentz
symmetry. It arises whenever the orbifold $\mathbbm{R}^6/S$ respects some
additional rotational symmetry beside $\theta =
\text{diag}(\mathrm{e}^{2\pi\I\, v_1},\mathrm{e}^{2\pi\I\, v_2},
\mathrm{e}^{2\pi\I\, v_3})$. For example, for a factorized orbifold, i.e.\ an
orbifold whose lattice $\Gamma$ is the direct product of three two--dimensional
lattices $\Gamma = \Gamma_1\times\Gamma_2\times\Gamma_3$, a rotation in the
sublattice $\Gamma_i$ by $\mathrm{e}^{2\pi\I\, v_i}$ is a symmetry of the theory. The
rotation by $v_i$ is of order $N_i$ (i.e.\ $N_i v_i \in\Z{}$) and results in a
$\Z{2 N_i}^R$ symmetry
\begin{equation}
 \sum_{r=1}^L -2R_r^i ~=~ 2 \mod 2N_i\;,
\end{equation}
where $R_r^i = q_{\text{sh},r}^i - \tilde{N}^i_r + \tilde{N}^{\bar\imath}_r$
with the oscillator numbers $\tilde{N}^i_r$ and $\tilde{N}^{\bar\imath}_r$
(see e.g.\ \cite{Buchmuller:2006ik} for their definition),
$q_{\text{sh},r}$ are the bosonic right--moving momenta and the factor $-2$
originates from the normalization such that fermions have a shifted $R$ charge
by $-1$. 

If the two--dimensional lattice $\Gamma_i$ has a higher symmetry than $N_i$
there is an additional string selection rule known as ``rule 4''. For example,
the $\SU{3}^3$ root lattice of a $\Z{3}$ orbifold allows for $\Z{6}$ sublattice
rotations. If all strings involved in a given interaction sit at the same fixed
point they feel the higher symmetry and the $R$ symmetry is enhanced to $\Z{4
N_i}^R$.

\end{enumerate}

\end{document}